\documentclass[12pt,a4paper,titlepage]{article}

\usepackage{indentfirst} %% Отсуп у первой строки раздела
\usepackage{lastpage} %% Количество страниц

\usepackage{amsmath} 
\usepackage{amssymb}
\usepackage{amsfonts}
\usepackage{amsthm}
\usepackage{bm} % Делает любые символы жирнными
\usepackage{icomma} % "Умная" запятая: $0,2$ --- число, $0, 2$ --- перечисление

\usepackage{extsizes} % Правильно меняет размер тескта
\usepackage{geometry} % Поля
\usepackage{setspace}
\usepackage{textcase}

\usepackage{cancel}

\usepackage[linkcolor=blue,colorlinks=true]{hyperref}

\newtheorem{lemma}{Lemma}
\newtheorem{theorem}{Theorem}

\theoremstyle{definition}
\newtheorem*{demo}{Proof}

\newcommand\tr{\operatorname{tr}}

\usepackage{epsfig}
\usepackage{graphicx}

\begin{document}
\begin{center}	
	
	\Large
	\textbf{Exact dynamics of moments and correlation functions for fermionic Poisson-type GKSL equations}
		
	\large 
	\textbf{Iu.A. Nosal}\footnote{Faculty of Physics, Lomonosov Moscow State University.},
	\textbf{A.E. Teretenkov}\footnote{Department of Mathematical Methods for Quantum Technologies,
		Steklov Mathematical Institute of Russian Academy of Sciences, Moscow, Russia.\\ E-mail:taemsu@mail.ru}
	
\end{center}

\footnotesize
Gorini-Kossakowski-Sudarshan-Lindblad equation of Poisson-type for the density matrix is considered. The Poisson jumps are assumed to be unitary operators with generators, which are quadratic in fermionic creation and annihilation operators.  The explicit dynamics of the density matrix moments and Markovian multi-time ordered correlation functions is obtained.\\

\textit{AMS classification:} 81S22, 82C31, 81Q05, 81Q80

\textit{Keywords:} GKSL equation, irreversible quantum dynamics, Poisson stochastic process, exact solution, fermions
\normalsize

\section{Introduction}

In this work we obtain both the fermionic analog of results of \cite{Teretenkov20} and  new result on multi-time ordered Markovian correlation functions. The latter one is also important due to modern interest to non-Markovian effects which manifest themselves only on the level of multi-time Markovian correlation functions rather than master equations  \cite{Gullo14}. As in \cite{Teretenkov20} we consider the equation for the density matrix
\begin{equation}\label{eq:mainEq}
\frac{d}{dt} \rho_t = \mathcal{L}(\rho_t), \qquad \mathcal{L}(\rho) = \sum_{k=1}^K \lambda_k (U_k \rho U_k^{\dagger} - \rho), \qquad \lambda_k >0,
\end{equation}
where $ U_k $ are unitary . Let us also note that the generator $ \mathcal{L} $  has the \textit{Gorini-Kossakowski-Sudarshan-Lindblad} (GKSL) form \cite{gorini1976completely, lindblad1976generators}
\begin{equation*}
\mathcal{L}(\rho) =  \sum_{k=1}^K \left(L_k \rho_t L_k^{\dagger} - \frac12 L_k^{\dagger} L_k\rho- \frac12 \rho L_k^{\dagger} L_k\right),
\end{equation*}
where $ L_k = \sqrt{\lambda_k} U_k $. 

In our case $ U_k $ are exponential of quadratic forms in fermionic creation and annihilation operators in the finite-dimensional Hilbert space. Such generators naturally arise in the case of averaging with respect to classical Poisson processes with intensities $ \lambda_k $ and unitary jumps $ U_k $ \cite{Kummerer87}, so we call them Poisson-type generators. For the infinite-dimensional Hilbert space such generators were discussed in \cite{Holevo96, Holevo98}. Let us note that Poisson processes and the correspondent quantum Markov equations arise in physical applications \cite{accardi2002quantum, vacchini2009quantum, Basharov2014, TrubBash2018}. Unitary evolution with the fermionic quadratic generators mentioned above was discussed in \cite{Fried1953, Ber86, Dodonov83}. Its bosonic counterpart was discussed in \cite{Fried1953, Ber86, Manko79, Manko87, dodonov2002nonclassical, dodonov2003theory, Cheb11, Cheb12}.

Now let us specify the exact mathematical formulation of our problem and main results. We use the notation which is similar to \cite{Ter17, Ter19}. We consider the finite-dimensional Hilbert space $\mathbb{C}^{2^n}$. In such a space one could  (see \cite[p. 407]{Takht11} for explicit formulae) define $ n $-pairs of fermionic creation and annihilation operators satisfying \textit{canonical anticommutation relations} (CARs): $ \{\hat{c}_i^{\dagger}, \hat{c}_j \} = \delta_{ij},  \{\hat{c}_i, \hat{c}_j\} = 0 $.  Let us define the $2n$-dimensional vector $\mathfrak{c} = (\hat{c}_1, \ldots, \hat{c}_n, \hat{c}_1^{\dagger}, \ldots, \hat{c}_n^{\dagger})^T$ of creation and annihilaiton operators. The  quadratic forms in such operators we denote by  $ \mathfrak{c}^T K \mathfrak{c} $, $ K \in \mathbb{C}^{2n \times 2n} $.  Define the $2n \times 2n$-dimensional matrix
\begin{equation*}
E = \biggl(
\begin{array}{cc}
0 & I_n \\ 
I_n & 0
\end{array} 
\biggr),
\end{equation*}
where $ I_n $ is the identity matrix from $ \mathbb{C}^{n \times n} $. Then CARs take the form $ \{f^T\mathfrak{c}, \mathfrak{c}^Tg \} = f^TEg$, $ f, g \in \mathbb{C}^{2n}$. We also define the $\sim$-conju\-ga\-tion of matrices by the formula
\begin{equation*}
\tilde{K} = E \overline{K} E, \qquad K \in \mathbb{C}^{2n \times 2n},
\end{equation*}
where the overline is an (elementwise) complex conjugation. 

\begin{theorem}\label{th:main}
	Let the density matrix $ \rho_t $ satisfy Eq.~\eqref{eq:mainEq}, where the unitary operators $ U_k$, $ k=1, \ldots, K $, are defined by the formulae $ U_k = e^{- \frac{i}{2} \mathfrak{c}^T H_k \mathfrak{c} } $, $ H_k = -H_k^T = -\tilde{H}_k \in \mathbb{C}^{2n \times 2 n} $,  then 
	
	1) Dynamics of the moments has the form
	\begin{equation}\label{eq:momDynam}
	\langle \otimes_{m=1}^M \mathfrak{c}   \rangle_t = e^{\sum_{k=1}^K\lambda_k (\otimes_{m=1}^M O_k  -  I_{(2n)^M}) t} \langle \otimes_{m=1}^M \mathfrak{c} \rangle_0, \qquad O_k = e^{-i E H_k},
	\end{equation}
	where the average is defined by the formula $ \langle \otimes_{m=1}^M \mathfrak{c} \rangle_t \equiv \mathrm{tr} \; (\rho_t \otimes_{m=1}^M \mathfrak{c}) $, $ I_{(2n)^m} $ is the identity matrix in $ \mathbb{C}^{2n} \otimes  \cdots \otimes \mathbb{C}^{2n}  = \mathbb{C}^{(2n)^m} $.
	
	2) If we denote $ L_{M, m} = \sum_{k=1}^K\lambda_k (\otimes_{r=1}^m O_k  -  I_{(2n)^m}) \otimes I_{(2n)^{M-m}}$ for $ m =1, \ldots, M  $, then the dynamics the Markovian multi-time ordered correlation functions has the form
	\begin{equation}\label{eq:corrEvol}
	\langle \mathfrak{c}(t_M) \otimes \ldots \otimes \mathfrak{c}(t_1)  \rangle = e^{L_{M,1} (t_M - t_{M-1})} \ldots e^{L_{M,M} t_1}	\langle \otimes_{m=1}^M \mathfrak{c} \rangle_0,
	\end{equation}
	where $ t_M \geqslant \ldots \geqslant t_1 \geqslant 0 $ and the tensor $ \langle \mathfrak{c}(t_M) \otimes \ldots \otimes \mathfrak{c}(t_1)  \rangle $ is defined by its elements
	\begin{equation}\label{eq:corrDef}
	\langle \mathfrak{c}_{j_M}(t_M) \ldots \mathfrak{c}_{j_1}(t_1)  \rangle \equiv \tr( \mathfrak{c}_{j_M} e^{\mathcal{L}(t_M - t_{M-1})}  \ldots \mathfrak{c}_{j_2} e^{\mathcal{L} (t_2 - t_1)}\mathfrak{c}_{j_1} e^{\mathcal{L} t_1}\rho_0), 
	\end{equation}
	where $  j_m =1, \ldots, 2n, $ for all $ m =1, \ldots, M $.
\end{theorem}

In definition \eqref{eq:corrEvol} for the Markovian multi-time ordered correlation functions we follow \cite{Gullo14}.

In particular, for the first and second moments we have
\begin{equation*}
\langle\mathfrak{c}\rangle_t = e^{\sum_{k=1}^K \lambda_k(O_k  -  I_{2n}) t} \langle\mathfrak{c}\rangle_0, \qquad \langle\mathfrak{c} \otimes \mathfrak{c}\rangle_t = e^{\sum_{k=1}^K \lambda_k (O_k \otimes O_k -  I_{4 n^2}) t} \langle \mathfrak{c} \otimes \mathfrak{c} \rangle_0
\end{equation*}
and for two-time correlation functions we have
\begin{equation*}
\langle\mathfrak{c}(t_2) \mathfrak{c}(t_1) \rangle = e^{L_{2,1} (t_2 - t_1)} e^{L_{2,2} t_1}  \langle \mathfrak{c} \otimes \mathfrak{c} \rangle_0.
\end{equation*}
Notr that $ \langle\mathfrak{c}(t) \mathfrak{c}(t) \rangle =  e^{L_{2,2} t}  \langle \mathfrak{c} \otimes \mathfrak{c} \rangle_0 = e^{\sum_{k=1}^K \lambda_k (O_k \otimes O_k -  I_{4 n^2}) t} \langle \mathfrak{c} \otimes \mathfrak{c} \rangle_0 = \langle\mathfrak{c} \otimes \mathfrak{c}\rangle_t  $.

\section{Dynamics in Heisenberg picture} 

In this section we prove Th.~\ref{th:main}. The main idea consists in transfer to the Heisenberg picture. To do it let us calculate the conjugate generator $ \mathcal{L}^*  $ defined by the relation
\begin{equation}\label{eq:defConjGen}
\mathrm{tr} \,  \hat{X} \mathcal{L}(\rho) = \mathrm{tr} \, \mathcal{L}^*  (\hat{X})\rho,
\end{equation}
for arbitrary matrices $ \rho, X \in \mathbb{C}^{2n \times 2n} $. We need lemma 1 from \cite{Ter17} in the case when $ A = i H $, $ B =0 $, which takes the following form.

\begin{lemma}
	\label{lem:orthTransform} 
	Let $ H = -H^T \in \mathbb{C}^{2 n \times 2n} $, then  $ e^{ \frac{i}{2} \mathfrak{c}^T H \mathfrak{c} }  \mathfrak{c} e^{- \frac{i}{2} \mathfrak{c}^T H \mathfrak{c} } = O \mathfrak{c}$, where $O = e^{-i E H} $.
\end{lemma}
Let us note that in accordance with lemma 4 from \cite{Ter17}, if $ \tilde{H} = - H$, then the matrix $ \frac12 \mathfrak{c}^T H \mathfrak{c} $ is self-adjoint. Thus, the operators $ U_k $ defined in Th.~\ref{th:main} are unitary indeed.

\begin{lemma}
	\label{lem:conjGen}
	Let $\mathcal{L} $ be defined by \eqref{eq:mainEq} with $ U_k = e^{- \frac{i}{2} \mathfrak{c}^T H_k \mathfrak{c} } $, $ H_k = \tilde{H}_k \in \mathbb{C}^{2n \times 2 n} $, and $ \mathcal{L}^*$ be defined by formula \eqref{eq:defConjGen}, then
	\begin{equation*}
	\mathcal{L}^*(\otimes_{m=1}^M \mathfrak{c}) =  \sum_{k=1}^K  \lambda_k (\otimes_{m=1}^M O_k - I_{(2n)^M}) \otimes_{m=1}^M \mathfrak{c} , \qquad  O_k = e^{-i E H_k}.
	\end{equation*}
\end{lemma}

\begin{demo}
	By the cyclic property of trace we have $ \tr \hat{X} (U_k \rho U_k^{\dagger} - \rho)  = \tr (U_k^{\dagger} \hat{X} U_k - \hat{X}) \rho  $. Hence, by Eq.~\eqref{eq:defConjGen} we obtain
	\begin{equation*}
	\mathcal{L}^* (\hat{X}) = \sum_{k=1}^K \lambda_k (U_k^{\dagger} \hat{X} U_k - \hat{X}).
	\end{equation*}
	Taking the elements of the tensor $ \otimes_{m=1}^M \mathfrak{c} $  as $ \hat{X} $ we obtain
	\begin{equation*}
	\mathcal{L}^*(\otimes_{m=1}^M \mathfrak{c} ) 
	=  \sum_{k=1}^K \lambda_k (U_k^{\dagger} (\otimes_{m=1}^M \mathfrak{c} ) U_k - \otimes_{m=1}^M \mathfrak{c} ) = \sum_{k=1}^K \lambda_k (  \otimes_{m=1}^M (U_k^{\dagger} \mathfrak{c} U_k)  - \otimes_{m=1}^M \mathfrak{c}). 
	\end{equation*}
	By lemma \ref{lem:orthTransform}, we have $ U_k^{\dagger} \mathfrak{c} U_k=  e^{\frac{i}{2} \mathfrak{c}^T H_k \mathfrak{c} } \mathfrak{c}  e^{- \frac{i}{2} \mathfrak{c}^T H_k \mathfrak{c} } =  e^{-i E H_k} \mathfrak{c} = O_k \mathfrak{c} $.
	Thus, we obtain
	\begin{equation*}
	\mathcal{L}^*(\otimes_{m=1}^M \mathfrak{c} ) 
	= \sum_{k=1}^K \lambda_k (  \otimes_{m=1}^M ( O_k \mathfrak{c} )  - \otimes_{m=1}^M \mathfrak{c}) =  \sum_{k=1}^K  \lambda_k (\otimes_{m=1}^M  O_k - I_{(2n)^M}) \otimes_{m=1}^M \mathfrak{c}. \quad \quad \qed
	\end{equation*}
\end{demo}

\noindent\textbf{Proof of Th.~\ref{th:main}}.

1) Taking into account lemma \ref{lem:conjGen} we obtain the Heisenberg evolution of the operators $ \otimes_{m=1}^M \mathfrak{c} $ in the following explicit form.
	\begin{equation}\label{eq:HeisEvol}
	e^{\mathcal{L}^* t}(\otimes_{m=1}^M \mathfrak{c}) = e^{ \sum_{k=1}^K  \lambda_k (\otimes_{m=1}^M  O_k - I_{(2n)^M}) t} \otimes_{m=1}^M \mathfrak{c}
	\end{equation}
	Then taking into account the definition of the average from the statement of Th.~\ref{th:main}  we have
	\begin{align*}
	 \langle \otimes_{m=1}^M \mathfrak{c} \rangle_t &\equiv \mathrm{tr} \; (\otimes_{m=1}^M \mathfrak{c} \rho_t) = \mathrm{tr} \; ( \otimes_{m=1}^M \mathfrak{c} e^{\mathcal{L} t}(\rho_0))   = \mathrm{tr} \; (e^{\mathcal{L}^* t}(\otimes_{m=1}^M \mathfrak{c})\rho_0) = \\
	 &= e^{ \sum_{k=1}^K  \lambda_k (\otimes_{m=1}^M  O_k - I_{(2n)^M}) t} \mathrm{tr} \; (\otimes_{m=1}^M \mathfrak{c} \rho_0) = e^{ \sum_{k=1}^K  \lambda_k (\otimes_{m=1}^M  O_k - I_{(2n)^M}) t} \langle \otimes_{m=1}^M \mathfrak{c} \rangle_0
	\end{align*}
	Thus, we obtain \eqref{eq:momDynam}.
	
2) As for the moments let us turn to Heisenberg evolution operators in definition \eqref{eq:corrDef}
\begin{equation*}
 \tr( \mathfrak{c}_{j_M} e^{\mathcal{L}(t_M - t_{M-1})}  \ldots \mathfrak{c}_{j_{2}} e^{\mathcal{L} (t_2 - t_1)}\mathfrak{c}_{j_1} e^{\mathcal{L} t_1}\rho_0) = \tr \rho_0 e^{\mathcal{L}^* t_1}((e^{\mathcal{L}^* (t_2 - t_1)}  ((\ldots e^{\mathcal{L}^* (t_M - t_{M-1})} \mathfrak{c}_{j_M}\ldots)\mathfrak{c}_{j_2}  ))\mathfrak{c}_{j_1}).
\end{equation*}
By formula \eqref{eq:HeisEvol} taking into account the definition of $ L_{1,1} $ we have
\begin{equation*}
e^{\mathcal{L}^* (t_M - t_{M-1})} \mathfrak{c}_{j_M} = (e^{L_{1,1}  (t_M - t_{M-1})} \mathfrak{c})_{j_M},
\end{equation*}
then
\begin{align*}
e^{\mathcal{L}^* (t_{M-1} - t_{M-2})}((e^{\mathcal{L}^* (t_M - t_{M-1})} \mathfrak{c}_{j_M}) \mathfrak{c}_{j_{M-1}}) &= e^{\mathcal{L}^* (t_{M-1} - t_{M-2})}( (e^{L_{1,1} (t_M - t_{M-1})} \mathfrak{c})_{j_M} \mathfrak{c}_{j_{M-1}}) =\\
= e^{\mathcal{L}^* (t_{M-1} - t_{M-2})}( (e^{L_{1,1} (t_M - t_{M-1})} \mathfrak{c}) \otimes \mathfrak{c})_{j_M j_{M-1}} &=  e^{\mathcal{L}^* (t_{M-1} - t_{M-2})}( e^{L_{2,1} (t_M - t_{M-1})} \mathfrak{c} \otimes \mathfrak{c})_{j_M j_{M-1}} =\\
= ( e^{L_{2,1} (t_M - t_{M-1})} e^{\mathcal{L}^* (t_{M-1} - t_{M-2})}(\mathfrak{c} \otimes \mathfrak{c}))_{j_M j_{M-1}} &= (e^{L_{2,1} (t_M - t_{M-1})} e^{L_{2,2} (t_{M-1} - t_{M-2})} \mathfrak{c} \otimes \mathfrak{c})_{j_M j_{M-1}}
\end{align*}
Analogously we have
\begin{equation*}
e^{\mathcal{L}^* t_1}((e^{\mathcal{L}^* (t_2 - t_1)}  ((\ldots e^{\mathcal{L}^* (t_M - t_{M-1})} \mathfrak{c}_{j_M}\ldots)\mathfrak{c}_{j_2}  ))\mathfrak{c}_{j_1}) = e^{L_{M,1} (t_M - t_{M-1})} \ldots e^{L_{M,M} t_1}	 \otimes_{m=1}^M \mathfrak{c}
\end{equation*}
By averaging with respect to the initial state $ \rho_0 $ we obtain \eqref{eq:corrEvol}. \qed 

\section{Conclusions}

In this work we have considered evolution for the density matrix in accordance with GKSL equation \eqref{eq:mainEq}. In part 1) of Th.~\ref{th:main} we have obtained the fermionic analog to Th.~1 from \cite{Teretenkov20}. We have also obtained multi-time ordered Markovian correlation functions, which is a  generalization of single-time formula \eqref{eq:momDynam} to the multi-time case. This is important due to the modern discussion of quantum Markovianity which necessarily (according to \cite{Gullo14}) leads to the very special form \eqref{eq:corrDef} for multi-time ordered correlation in addition to the GKSL from of the master equations. The explicit expression for these correlation functions in our case is presented in part 2) of Th.~\ref{th:main}. The study of Markovian and non-Markovian effects are important now due to rising interest to the open quantum systems, the range of approaches to which is becoming wider and wider now \cite{Trushechkin19, Trushechkin19a, Luchnikov19, Teretenkov19a, Teretenkov19b}. A possible direction of future development consists in calculation of more general multi-time observables, e.g. unordered correlation functions in 2D echo-spectroscopy \cite{Plenio13}.

\end{document}